\documentclass[pdftex,twocolumn,epjc3]{svjour3}          

\RequirePackage[T1]{fontenc}

\smartqed  

\RequirePackage{graphicx}
\RequirePackage{mathptmx}      
\RequirePackage[numbers,sort&compress]{natbib}
\RequirePackage[colorlinks,citecolor=blue,urlcolor=blue,linkcolor=blue]{hyperref}

\journalname{Eur. Phys. J. C}

\usepackage{dcolumn}
\usepackage{color}
\usepackage{booktabs}
\usepackage{xspace}
\usepackage{graphicx}
\usepackage{amsmath}
\usepackage{amssymb}
\usepackage[figuresright]{rotating}
\usepackage{subfigure}
\usepackage{lineno}
\usepackage{xspace}

\graphicspath{
 {figure/}
}

\def \be  {\begin{equation}}
\def \ee  {\end{equation}}
\def \ee  {\end{equation}}
\def \bea {\begin{eqnarray}}
\def \eea {\end{eqnarray}}

\newcommand{\pp}{pp\xspace}
\newcommand{\ppb}{p--Pb\xspace}

\newcommand{\pbpb}{Pb--Pb\xspace}
\newcommand{\AAA}{AA\xspace}

\newcommand{\RAA}{\ensuremath{{R_{\rm AA}}}\xspace}

\newcommand{\pt}{\ensuremath{{p_{\rm T}}}\xspace}

\newcommand{\ccbar}{\ensuremath{c\overline{c}}\xspace}
\newcommand{\jpsi}{\ensuremath{J/\psi}\xspace}
\newcommand{\DZ}{\ensuremath{D^0}\xspace}
\newcommand{\DZbar}{\ensuremath{\overline{D}^0}\xspace}

\newcommand{\snnt}[1]{\ensuremath{\sqrt{s_{\rm NN}} = #1 \text{\,TeV}}\xspace}

\begin{document}
\title{Probing the Gluon Plasma with Charm Balance Functions}

\author{  
Sumit Basu\thanksref{addr1} \and
Peter Christiansen\thanksref{addr1,e1} \and
Alice Ohlson\thanksref{addr1} \and
David Silvermyr\thanksref{addr1}
}
\thankstext{e1}{e-mail: {\em peter.christiansen@hep.lu.se} (Corresponding author.)}
\institute{
Lund University, Department of Physics, Division of Particle Physics, Box 118, SE-221 00, Lund, Sweden\label{addr1} 
}

\date{Received: date / Revised version: date}
\maketitle
\begin{abstract}
  Recent theoretical explanations for how hydro\-dynamic-like flow can build
  up quickly in small collision systems (hydrodynamization) has led to a
  microscopic picture of flow building up in a gluon-dominated phase before
  chemical equilibrium between quarks and gluons has been attained. The goal
  of this contribution to Offshell-2021 is to explore consequence of assuming
  a long-lived gluon-dominated phase, which we shall denote a gluon plasma
  (GP). As these consequences are naturally enhanced in a large systems, we
  assume and explore the extreme scenario in which a GP would be created in
  \AAA collisions and exist for significant time before the formation of a
  chemically-equilibrated quark-gluon plasma (QGP). The GP and its formation
  would be impossible to probe with light-quark hadrons, which are first
  produced later in this scenario. As charm quarks are produced early in the
  collision, they can circumvent the limitations of light quarks and we
  propose charm balance functions as an effective tool to test this idea and
  constrain the dynamics of the GP.
\end{abstract}

\section{Introduction}

One of the biggest open questions in the study of heavy-ion collisions is how
the initial interactions lead to the formation of a strongly-interacting
medium. When the perfect liquid was discovered at
RHIC~\cite{Adams:2005dq,Adcox:2004mh,Busza:2018rrf} it was conjectured that
hydrodynamic behavior implied the formation of a thermal medium. With the discovery of
flow in small collision
systems~\cite{Khachatryan:2010gv,Abelev:2012ola,Nagle:2018nvi}, such as \pp
and \ppb collisions, at the LHC, the question of thermalization has become a hot
topic since it is hard to understand how a small system can have time to
thermalize. It appears that the concept of hydrodynamization must be part of
the solution to this open issue~\cite{Romatschke:2017ejr}. Hydrodynamization
is the phenomena that a system out of equilibrium can sometimes still be
described by hydrodynamics. Recently, a microscopic picture of
hydrodynamization has been developed based on kinetic theory, see
e.g. Ref.~\cite{Kurkela:2018xxd}. In these calculations, the initial state is
completely dominated by gluons and it is in this state that the system
hydronamizes. Light quarks are eventually produced and the system then first
attains chemical and then later thermal equilibration
(see Ref.~\cite{Kurkela:2018xxd} for a detailed discussion). This picture is very
similar to the 30-year-old idea of ``Hot
Glue''~\cite{Shuryak:1992wc} and more recent versions like ``Undersaturated
QGP''~\cite{Stocker:2015nka}, which are all driven by the expectation that the
initial state will be gluon-dominated due to the fact that the magnitude of
the color charge of the gluons is approximately twice that of the quarks.  As a consequence, the
initial projectile/target wave functions become gluon dominated at large beam
energies and at the same time gluons are more likely to scatter or be
produced in an annihilation process both in initial and final state
interactions. While there has been some previous work on this idea, we stress that
this is not the standard picture of QGP formation.\\

In this paper, we
resurrect these speculative ideas and reexamine them. We think it is timely to do so because of the recent realization that hydrodynamic flow does not
require thermalization. This realization means that the system can exhibit properties traditionally associated with a QGP medium -- namely, collective flow (in kinetic equilibrium) -- but
not yet be in chemical or thermal equilibrium, which is unlike the standard QGP picture of the last 20 years. If such a system is indeed
produced, and it has QGP-like properties but is gluon-dominated, we propose to
denote it the gluon plasma (GP) and reserve QGP for the chemically-equilibrated system. The main goal of this paper is to point out that the
possible existence of a GP phase leads to a different view of classic QGP
observations and to identify how we can get further insights into this idea
experimentally.\\  

Before we go on, we stress that in a weakly-coupled
calculation (e.g. as discussed in Ref.~\cite{Kurkela:2018xxd}) one would
\emph{not} expect a significant impact of a GP/``Hot Glue'' phase. However,
the weak coupling assumption is questionable~\cite{Shuryak:2019ydl} and
likely enforced by the lack of theoretical tools to do the strongly-coupled
microscopic calculation. We therefore here \emph{assume} and explore a more
extreme strongly-coupled scenario where the GP phase would last for a
significant time and dominate the initial system. In this way, we can point
out both the strengths of such a hypothetical picture and how we can test it
experimentally.

As we shall see, the existence of a significant GP phase could provide novel insights into
some open theoretical issues, but also presents experimental challenges: if true, it
would imply that light quarks are produced late in the collision evolution and so light
hadrons will likely be less sensitive to early-time dynamics. For this reason,
we propose to use low \pt charm quarks to study the dynamics of the
hydrodynamization process. The advantage of using charm quarks is that:
\begin{itemize}
\item they are dominantly produced early in the collision with rates that
  are calculable in perturbative QCD (pQCD)
\item they are the most abundantly produced heavy quarks
\item they are known to interact with the medium and exhibit collective
  flow~\cite{ALICE:2018bdo}
\end{itemize}
For completeness, we note that the charm quantum number, $C$, is conserved by
strong and electromagnetic interactions while it can be violated in weak
decays. This means that charm quarks are essentially always created in pairs
(\ccbar). As weak decays predominantly occur over longer timescales than the hadronization process, it is still
possible to reconstruct the original charm hadrons and determine their charm content ($c$
or $\overline{c}$)\footnote{For neutral $D$ mesons, one can have oscillations,
  e.g., \DZ--\DZbar oscillations, which violate $C$, but this effect is
  essentially negligible.}.

This paper is organized as follows: Sec.~\ref{sec:gp} describes the
difference between the GP and the QGP and discusses general concepts used in
the rest of the paper. Section~\ref{sec:charm:anticharm} focuses on the idea to
study the charm-anticharm balance functions. In Sec.~\ref{sec:jpsi} we discuss
correlations between charm hadrons and ``regenerated'' \jpsi and finally in
Sec.~\ref{sec:conclusions} we summarize the main conclusions of the paper.

\section{The Gluon Plasma vs the Quark-Gluon Plasma}
\label{sec:gp}

In this section we discuss the idea that heavy-ion collisions lead to
the early formation of a GP and it is only later, after significant flow has
been built up in the system, that a QGP is formed. The goal is to try to
convince the reader that there are some existing results that can be more easily
understood if the initial system formed is a GP.\\  

The essential differences that a GP introduces, with respect to the standard QGP picture, are:
\begin{itemize}
\item an absence of light quarks at early times when the system is the hottest
\item that light quarks are created flowing.
\end{itemize}
Two heavy-ion observables that are the most sensitive to early light quark
dynamics are thermal photon production and the chiral magnetic effect.\\

Thermal photons\footnote{Experimentally, one measures direct (non-decay)
  photons, which are expected to be a mix of prompt photons (dominant at high \pt) and
  thermal photons (dominant at low \pt in central \AAA collisions).} have been
used to study the thermal properties of the initial
state~\cite{Adare:2009qk,Adam:2015lda}. It is clear that photons have no
direct coupling to gluons, which are electrically neutral, and so one would
naively expect that the dominant source of thermal photons will be thermal
radiation from light quarks. If a GP precedes the QGP phase, then one would
expect thermal photons to have a lower temperature because they would only be
produced later in the evolution of the system after the ``fireball'' has
already expanded and cooled, \emph{and} that they will be produced from a flowing system and thus exhibit collective flow behavior. While it is
unclear if the measured photon spectrum is cooler than expected, it was a huge
surprise that the thermal photons exhibited azimuthal anisotropic
flow in heavy-ion collisions~\cite{Adare:2011zr}. This puzzle is still unresolved and the existence of
a GP phase may be a key to resolve this question. These ideas have already been
explored in a similar context in Refs.~\cite{Liu:2012ax,Monnai:2014kqa}.\\

The observation of the chiral magnetic effect (CME) would be a signature of
parity violation in the strong interaction.  The CME is postulated to arise if
there are topological regions within the QGP with non-zero net chirality
(i.e. an excess or depletion of left- or right-handed quarks).  Such domains
may result in an experimentally-observable asymmetry in charged particle
emission because they would exist in the presence of a huge magnetic field which arises at very early times from the angular
momentum of the non-interacting spectators in non-central heavy-ion collisions.  The electric dipole, which would
be created by light quarks aligning their spins with this magnetic field,
would result in a charge splitting in the direction of the magnetic field. For
a review, see Refs.~\cite{Kharzeev:2013ffa, Kharzeev:2015kna}.\\

While the CME
has been observed in semimetals (e.g. ZrTe$_5$ ~\cite{Li:2014bha}), the
experimental search for such charge splitting in heavy-ion collisions has been
challenging due to the presence of background effects like collective flow and
local charge conservation which can mimic CME-like signals.  Recent efforts to
measure the CME and estimate the level of background contributions indicate
that \emph{at most} only a small fraction of the observed CME-like signals may
be attributable to chiral effects~\cite{Acharya:2017fau, Khachatryan:2016got,
  Acharya:2020rlz}.  This lack of a significant observed CME
signal~\cite{ALICE:2017sss,STAR:2021mii} may also be consistent with the
picture that the early stages of the heavy-ion collision system are a GP
phase, where quarks have not yet been created. We are not aware of anyone else
pointing out this possibility but we think the GP solution to the CME
puzzle is very much in the spirit of the rest of this paper because it is
not that there is no CME, it is rather that there are no quarks to probe it!\\

\begin{figure}[tbp]
\centering
\includegraphics[width=0.9\columnwidth]{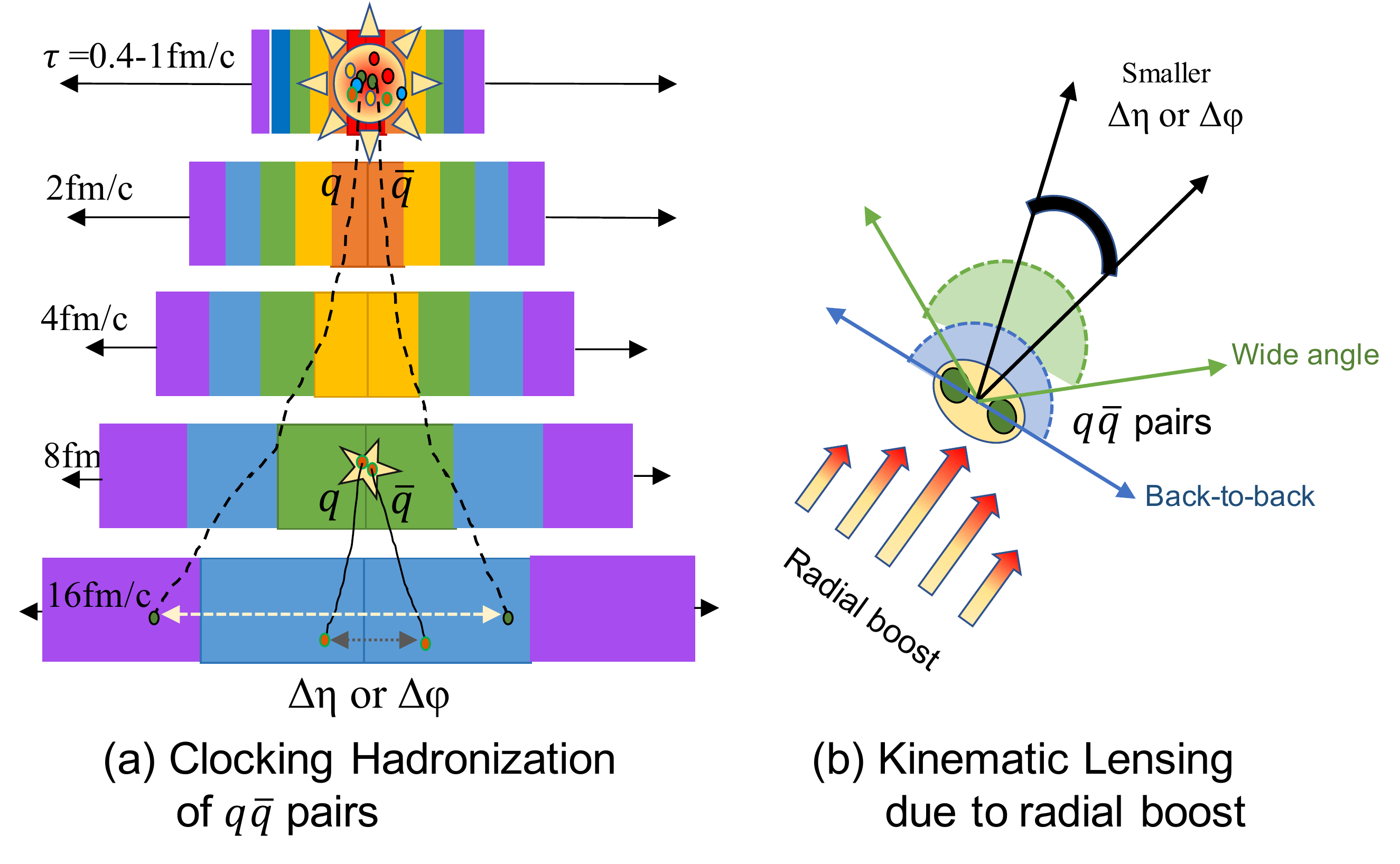}
\caption{Illustration of the physics affecting the width of the balance
  function. (a) Charges produced early in the collision will be further
  separated in momentum space than charges produced later (Inspired
  by Ref.~\cite{Bass:2000az}). (b) Radial flow can lead to a narrowing of the
  balance function due to kinematic lensing or focusing (Inspired by Fig.~9
  in Ref.~\cite{Nahrgang:2013saa}).}
\label{fig:balance_illu}
\end{figure}

Another result that is very important for the discussion in the next section
is related to balance functions~\cite{Bass:2000az,Jeon:2001ue,Pratt:2015jsa,
  Bialas200431}. Balance functions are used to measure how far in
momentum-space one has to go to balance quantum numbers, e.g., the charge
balance function measures where in momentum-space the electric charge is
balanced. Figure~\ref{fig:balance_illu} gives an illustration of what physics
affects the width of the charge balance functions:
\begin{itemize}
\item Clocking Hadronization: the width is broader (narrower) if the quarks that make up the
  hadrons are produced early (late), as they have more (less) time to diffuse ~\cite{Bass:2000az}.
\item Kinematic Lensing: the angle between pairs of particles created at rest
  in the comoving frame will be narrowed (focused) when boosted to the lab frame ~\cite{Abelev:2013csa}.
\end{itemize}

\begin{figure}[tbp]
\centering
\includegraphics[width=0.8\columnwidth]{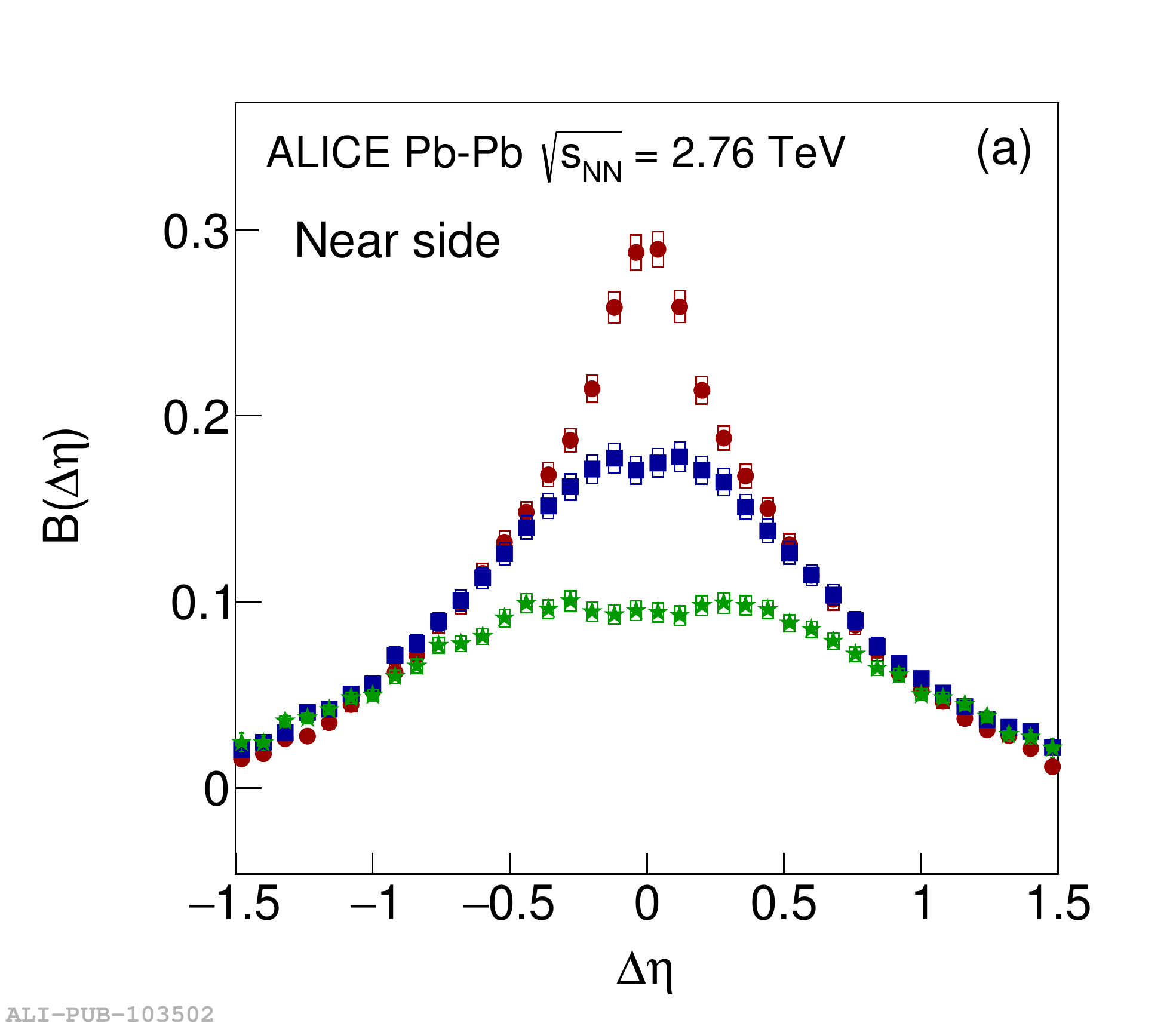}
\includegraphics[width=0.8\columnwidth]{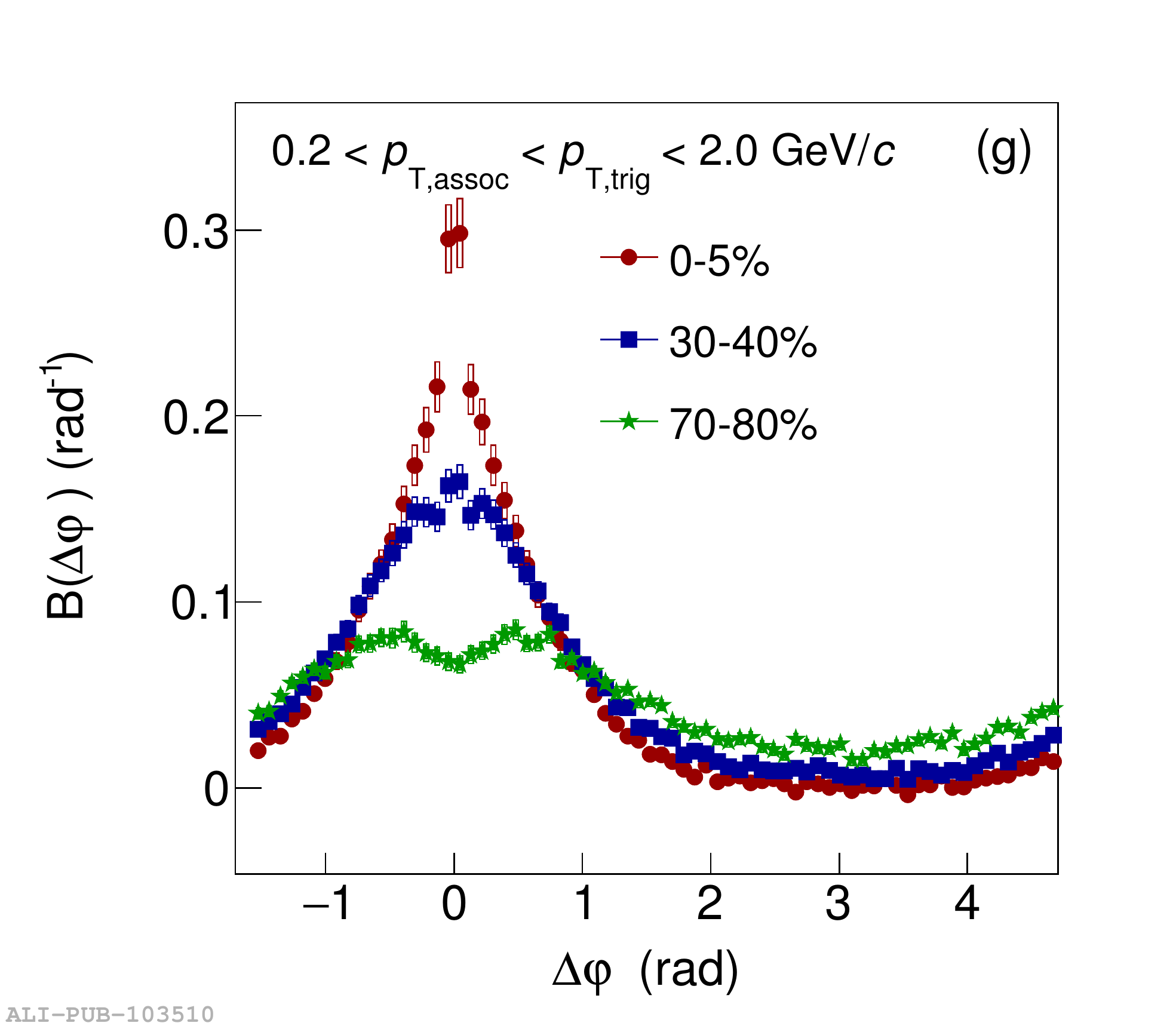}
\caption{Top: The balance functions in \snnt{2.76} \pbpb collisions as a
  function of the pseudorapidity separation between two charged particles on
  the near side (for relative azimuthal separations $-\pi/2 < \Delta\varphi <
  \pi/2$). One observes a clear narrowing in central collisions. Taken
  from Ref.~\cite{Adam:2015gda}. Bottom: Similar to the top plot but for
  separation in azimuthal angle.}
\label{fig:balance_intro}
\end{figure}

The top and bottom panels of Figure~\ref{fig:balance_intro} present
experimental results for the charge balance function in \snnt{2.76} \pbpb
collisions. It is observed that for more central, longer-lived collisions, the
balance function is narrower. In the QGP picture, the interpretation is that,
even if quarks are produced early, they stay in thermal equilibrium (pairs are
created and annihilated) until hadronization and so the charge balance is produced
late when the system is flowing. In this picture, the balance would be
produced at a similar $T_c$ for all centralities (Late Clocking Hadronization)
and the main difference is then due to the different amount of Kinematic
Lensing. In the GP picture the balance of light quarks is naturally created
late when the system is already flowing and so one would reach a similar
conclusion but without the requirement that quarks are balanced down to $T_c$,
which seems to be in contrast with the widespread idea of a grand canonical
ensemble.  \\ It might seem that the narrowing of the charge balance function
is of little importance but one of the main signatures of a QGP must be
deconfinement, i.e. that pairs of quarks created together at early times will
hadronize far apart at later times. This type of deconfinement is exactly
what differentiates the QGP from confined descriptions, such as the Lund string
model~\cite{Andersson:1983ia,Sjostrand:2007gs,Sjostrand:2014zea}. The irony of
the GP scenario would be that the initial state is made up of deconfined gluons but
that we cannot easily measure this deconfinement, since we are predominantly sensitive to the quarks
(hadrons), which are produced late and kinematically focused. In
Sec.~\ref{sec:charm:anticharm} we therefore extend the balance
function to charm-anticharm quarks which are produced early in the collision
and therefore should behave completely differently from the light hadrons
shown in Fig.~\ref{fig:balance_intro}.\\

Finally we touch upon two ideas that we hope can inspire the
reader.
\begin{enumerate}
    \item If a GP phase exists, it will not be described by standard lattice
QCD (LQCD) calculations as it is out of chemical equilibrium and it could
imply that the screening of the \ccbar potential is not well described by
LQCD. 
\item The focus in the following will be dominantly on LHC
physics. However, we note that there could be a difference between the
relative importance of the GP at the LHC, RHIC, and the SPS, as one would expect the
initial states to be more quark-dominated at lower beam energies.
\end{enumerate}

\section{Charm-anticharm balance functions}
\label{sec:charm:anticharm}

Heavy quarks have played an important role in studies of the QGP for a long
time. Initially, quarkonia were mainly studied due to the predicted and
observed melting in a deconfined
medium~\cite{Matsui:1986dk,Abreu:2000ni,Adare:2006ns}. However, in particular
at LHC energies, the charm mesons have become a hot topic because they are
well-defined probes of the produced medium: the initial charm quark production
can be precisely estimated in heavy-ion collisions, their mass is large with
respect to the temperature of the produced medium and a large fraction of the
charm quarks are non-relativistic. For a recent overview, see for example
Ref.~\cite{Dong:2019byy}.\\

So far, experimental measurements have primarily focused on the spectra of charm hadrons, but these provide only coarse
information about final state interactions and one must now go on to study
correlations. In the GP picture, correlating charm hadrons with light-flavor
hadrons (as has been done in both correlation and flow measurements) does not
probe the build-up of flow signals because the light quarks are produced after
the system is already flowing. Therefore we think that in order to access the earliest stages of the collision, one must now go on to study
correlations between charm hadrons. \\ For charm-(anti)charm correlations, we
first point out that one can in principle calculate the initial momentum
correlations between \ccbar pairs using pQCD, see
e.g. Ref.~\cite{Vogt:2018oje}. The initial momentum correlations will be the same
in peripheral and central heavy-ion collisions because the microscopic
scattering process is the same, and charm quarks from independent scatterings
will be uncorrelated in momentum space. The production rate will be scaled up
to account for the nuclear overlap geometry. We also mention that
phenomenological studies of charm-charm correlation functions have previously
been done for flow~\cite{Akamatsu:2009ya},
thermalization~\cite{Zhu:2006er,Tsiledakis:2009qh}, jet
quenching~\cite{Nahrgang:2013saa}, and even for the effects of pre-equilibrium
physics~\cite{Ruggieri:2018ies}. The idea to study charm-charm correlation
functions (and even balance functions) is therefore not new and even the idea
to use it as a probe of early-time physics can be found in some of these
papers. What is new in this paper is the idea of the GP and how to probe it
with charm balance functions, the specific ideas of Sec.~\ref{sec:jpsi}, but
maybe most importantly we hope to convince the reader that these types of
correlation measurements, that are extremely challenging from an experimental
point of view, are no longer ``nice to have'' but really something
we ``need to have'' as a community.

\subsection{What type of correlations should be measured?}

We propose to measure the correlation between the charm and the anticharm
quarks produced in the \emph{same} hard process. It is an extension of ideas
described in Ref.~\cite{Adolfsson:2020dhm} for strange-antistrange
correlations and we propose to use the same analysis method as described in
Refs.~\cite{Adolfsson,Adolfsson:2020sxz}. \\ First of all, one cannot measure
the charm quarks directly and so one has to measure the charm hadrons, e.g.,
$\DZ(c\overline{u})$ and
$\DZbar(\overline{c}u)$. In the model calculations shown in this
paper we will generally just sum over all charm hadrons and all anticharm
hadrons for the results but this of course cannot be done
experimentally. Secondly, one does not know if a given charm hadron and a particular anticharm
hadron are from the same initial process. To access the directly-correlated
pairs we propose to do as in Refs.~\cite{Adolfsson,Adolfsson:2020sxz} and subtract
charm-charm and anticharm-anticharm correlations from charm-anticharm
correlations. In this way, and if we properly normalize the correlation
functions, we are left with the direct correlations, which is the balance
function as described in Sec.~\ref{sec:gp}.  We will therefore be able to
access the early-time dynamics (see Fig.~\ref{fig:balance_intro}, top panel)
by measuring how the balance function in heavy-ion collisions is modified with the respect to the one
calculable (measurable) in pQCD (\pp collisions). We note that there are
slightly different ways to define and normalize the balance function
mathematically, see for
example Refs.~\cite{Bass:2000az,Abelev:2013csa,Adam:2015gda,Basu:2020ldt}, but that
they are in essence the same and boil down to measuring experimentally where
in momentum-space the charm quantum number is balanced.

\begin{figure}[tbp]
\centering
\includegraphics[width=\columnwidth]{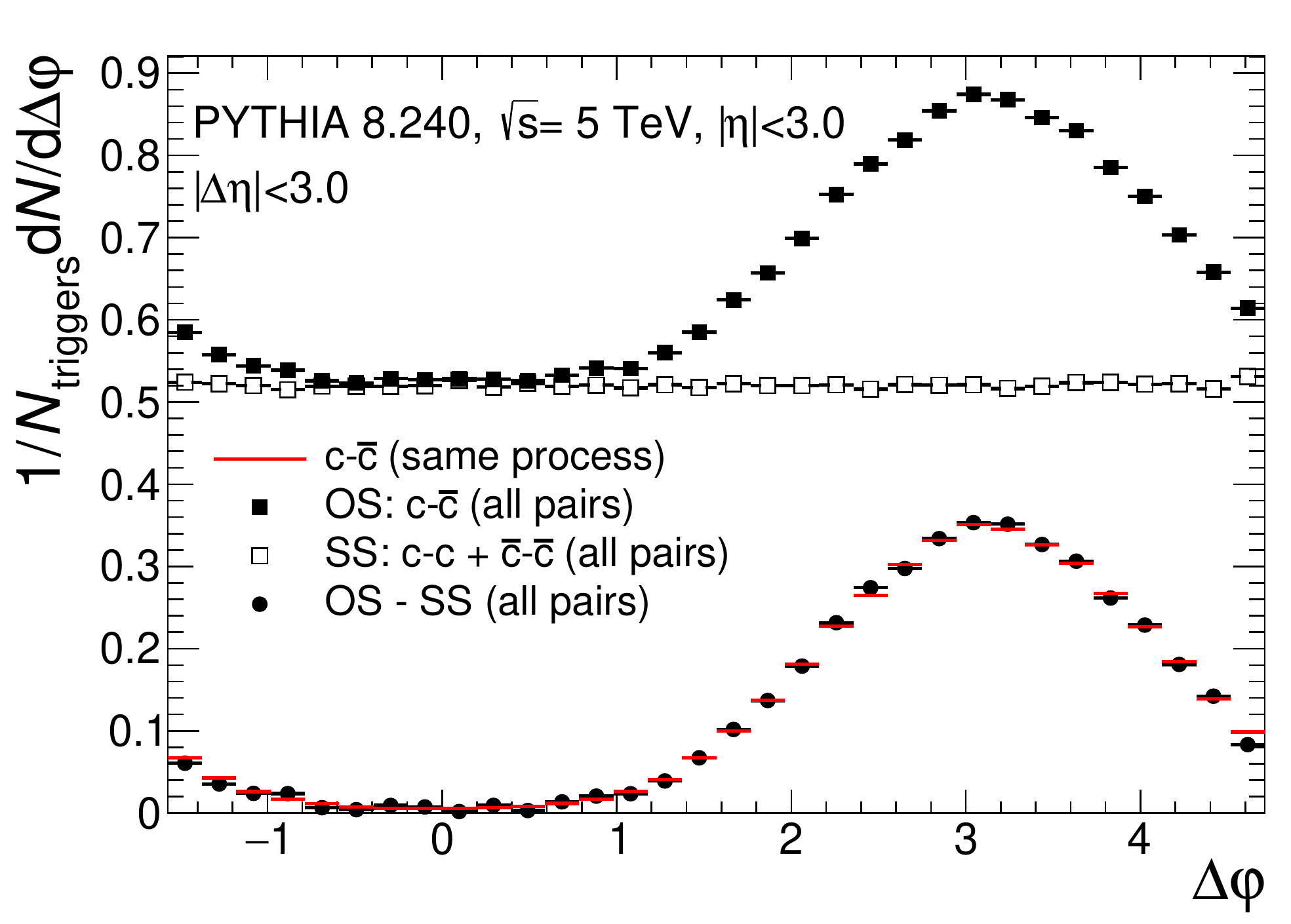}
\includegraphics[width=\columnwidth]{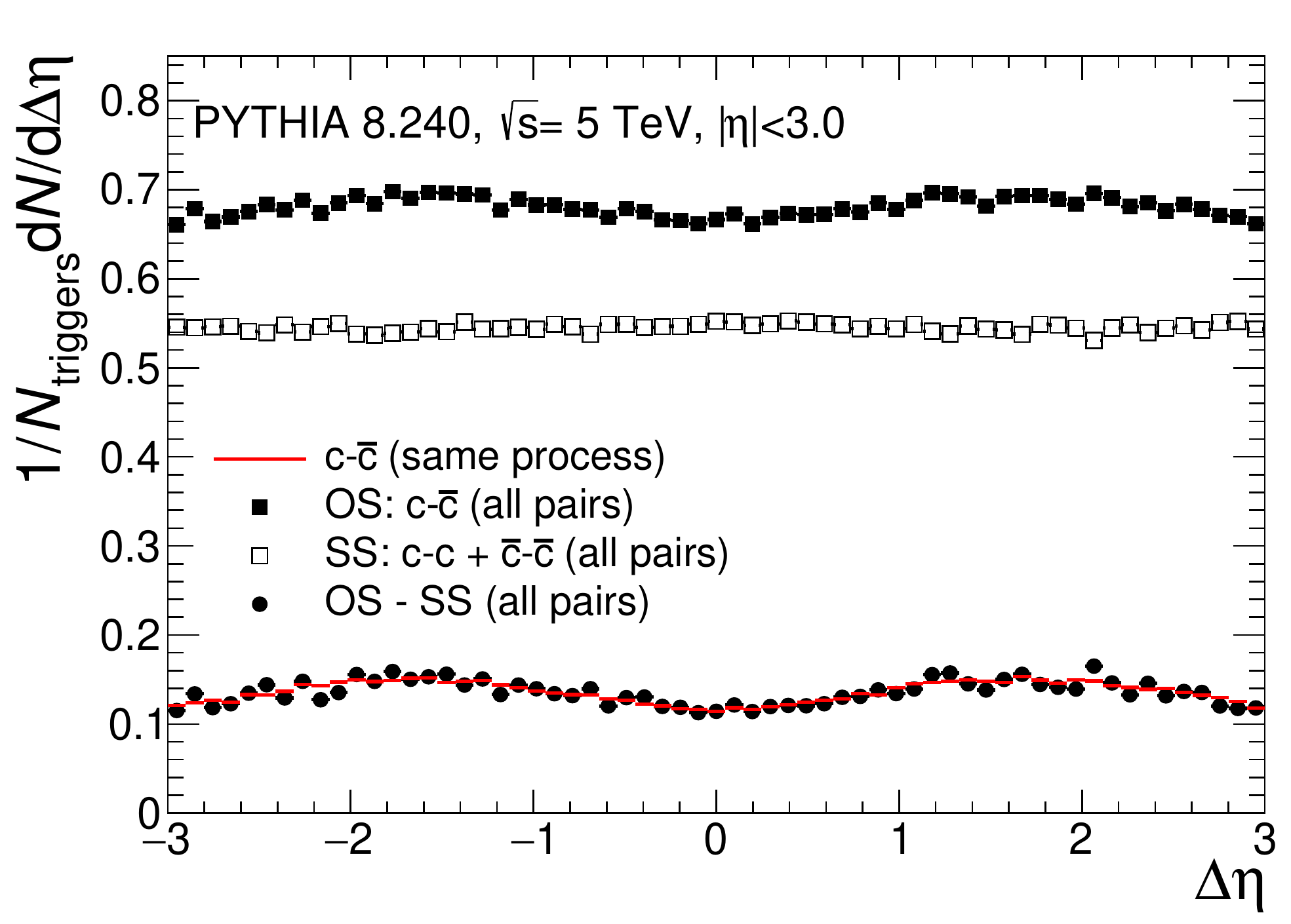}
\caption{Illustration of how one experimentally can extract the correlation
  between correlated charm and anticharm quarks produced in the same hard
  process (red line). The experimental observable (OS-SS) is constructed by
  subtracting the same sign (SS) per trigger yield from the opposite sign (OS)
  per trigger yield. Top: azimuthal correlations. Bottom: Pseudorapidity
  correlations. Note that results are shown for all charm hadrons and that $c$
  and $\overline{c}$ only refers to the charm content of the hadrons.}
\label{fig:balance_function}
\end{figure}

To illustrate the idea behind the charm balance function, in particular the
subtraction scheme, we have carried out a simulation study using
PYTHIA~\cite{Sjostrand:2006za,Sjostrand:2014zea}. We have generated events
where we have forced the hardest scattering to produce a pair of \ccbar quarks. In some cases more than one pair of charm
hadrons have been produced (\ccbar can also be produced in sub-leading hard
processes). These events have been rejected to ensure that we always have only
one pair of correlated charm quarks, but we note that for these events there
was no angular correlation between the leading and sub-leading pairs of charm
hadrons. Figure~\ref{fig:balance_function} shows the correlation between the
charm and anticharm hadrons produced in the same hard scattering (red
line). Here we point out that in
Fig.~\ref{fig:balance_function}, the charm and anticharm hadrons are obviously
back-to-back in azimuthal angle, but LHCb has actually measured these
correlations in \pp collisions and found a significant near-side
peak~\cite{LHCb:2012aiv}. To get such a near-side peak one would need
significant NLO corrections as discussed in Ref.~\cite{Vogt:2018oje}. So one
should be aware that the real correlations could be very different from the
ones obtained at LO in PYTHIA. In fact, one could wonder if it would be possible to
experimentally constrain NLO contributions to the charm production process using this type of
correlation measurement. We note that since the correlations in \pp collisions are mainly
considered here as a reference for \AAA measurements, this has no impact on
the arguments that follow.

One challenge in obtaining the correlation function shown in Fig.~\ref{fig:balance_function} is that we will
experimentally measure both the correlated and uncorrelated pairs.  Here we show our idea for
how to remove the uncorrelated pairs experimentally: To construct the correlation function,
we mix each event with 6 other random events and measure that same sign (SS) and
opposite sign (OS) per-trigger yields. Now, by subtracting the SS from the OS
yields one can in fact recover the directly correlated yield between the charm
and the anticharm quarks, which we will call the charm-anticharm balance
function in the following. \\

We have not included detailed model predictions for what one will be able to
measure using the charm-anticharm balance function. The concept of the GP is a
novel idea and it is not yet implemented in any standard generators, but we
expect that one should be able to get some idea about this in kinetic theory,
e.g., using the formalism of Ref.~\cite{Kurkela:2018xxd}. We also note that an
out-of-equilibrium calculation, using the Glasma formalism, was presented
in Ref.~\cite{Ruggieri:2018ies}. At the qualitative level, we predict that one should observe something very different from what was measured for light quarks, cf.\ Fig~\ref{fig:balance_intro}. In fact, in simulations done for jet quenching, which is modeled as an
interaction between the medium and the pair of charm quarks
in Ref.~\cite{Nahrgang:2013saa}, then a broadening of the balance
function is observed. We think that the observation of a broadening at low \pt would
establish that the physics one can extract from the charm-anticharm balance
function would be of a different nature from what the light quarks probe.\\

Finally, we reiterate the main ideas here:
\begin{itemize}
\item Charm is dominantly produced in hard scatterings in the initial stages of the collision, which means
  that one can establish a precise charm balance function reference in small
  systems
\item It is known that charm interacts with the medium as it builds up flow
\item We propose to measure how the charm balance function is modified as the
  system size is increased
\end{itemize}
Depending on what kind of statistics can be achieved, one could even imagine
combining this with event-shape engineering~\cite{Schukraft:2012ah} to ensure
that all initial states have similar geometries while one
mainly varies the system lifetime.
  
\section{Correlations between $\mathbf{\jpsi}$ and charmed hadrons}
\label{sec:jpsi}

The production of charmonium in general, and \jpsi in particular, is one of
the most studied signatures in heavy-ion collisions, since the production is
expected to be affected by color screening in the dense QGP
state~\cite{Matsui:1986dk}. At the LHC, it is observed that the \jpsi nuclear
modification factor \RAA\footnote{\RAA quantifies the modification of hard
  processes in \AAA collisions with respect to an independent superposition of \pp collisions, taking into
  account the enhanced production due to the nuclear geometry; see
  e.g. Ref.~\cite{Dong:2019byy} for details.} is larger than what has been measured
at RHIC at low \pt~\cite{Abelev:2012rv,Abelev:2013ila}. One normally expects that at higher
beam energies, the plasma is hotter and therefore the screening/suppression should be more significant. To explain this difference it has been proposed that \jpsi
``regeneration'' plays a large role at LHC: if charm is copiously produced it
can occur that a charm quark produced in one hard scattering can combine with
an anticharm quark produced in another \emph{independent} hard scattering to
form a \jpsi.  Since charm is more abundantly produced at LHC energies than at RHIC, the larger effect of regeneration, on top of the expected suppression, leads to a higher \RAA at the LHC~\cite{BraunMunzinger:2000px}. \\

If \jpsi regenerates at the LHC, it would be a direct signature that
charm/anticharm quarks are deconfined in the plasma. Therefore one should
consider if there are alternative explanations. A challenge in interpreting \RAA measurements is
that it depends on the beam energy, since the \pp reference spectrum will
typically harden as the beam energy increases. The same modification of the
\AAA spectrum at two different beam energies can therefore lead to two
different values of \RAA. An alternative reason for the change in the \jpsi \RAA from RHIC to the LHC could
be if GP effects play a larger role at the LHC and result in less screening of
the charmonium potential. For this reason, we propose in the following to
search for correlations between \jpsi from regeneration and other charm
hadrons.\\

\begin{figure}[tbp]
\centering
\includegraphics[width=1.0\columnwidth]{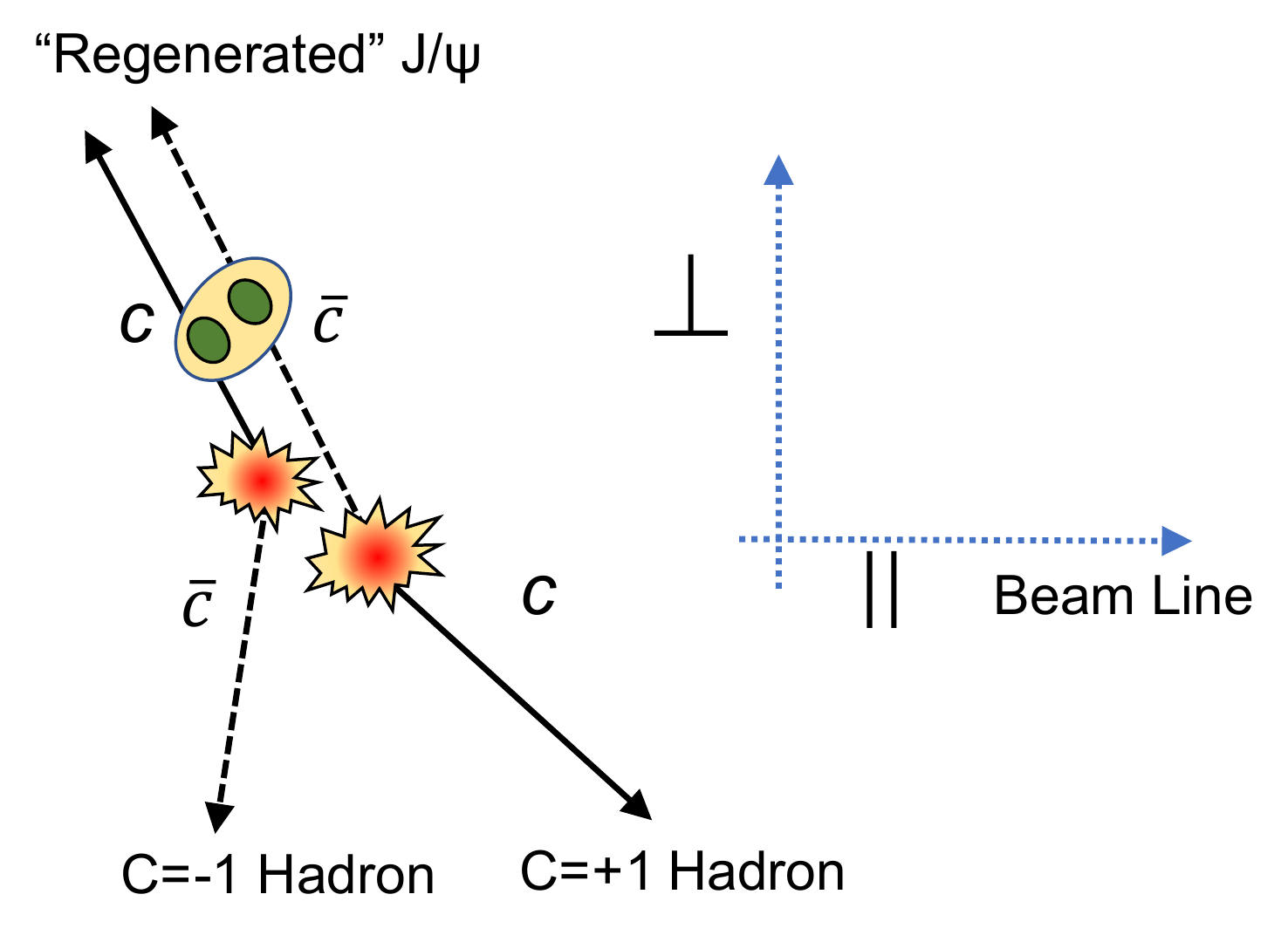}
\caption{Illustration of the \jpsi ``regeneration'' and how it leads to
  correlations between the \jpsi and the ``associated'' charm hadrons.}
\label{fig:jpsi_corr}
\end{figure}

Figure~\ref{fig:jpsi_corr} illustrates how \jpsi regeneration leads to
correlations between the \jpsi and charm hadrons. What is important to stress
here is that one normally does not expect such correlations to be present as
the \jpsi is expected to be directly produced. LHCb has in fact measured
correlations between \jpsi and D mesons and this correlation is approximately
flat in azimuthal angle~\cite{LHCb:2012aiv}, indicating that there are little
or no direct correlations between these hadrons under normal circumstances. We
therefore propose to study \jpsi--charm-hadron correlations to validate that
indeed \jpsi regeneration is the explanation for the increase in \RAA at low
\pt. \\

\begin{figure}[tbp]
\centering
\includegraphics[width=1.0\columnwidth]{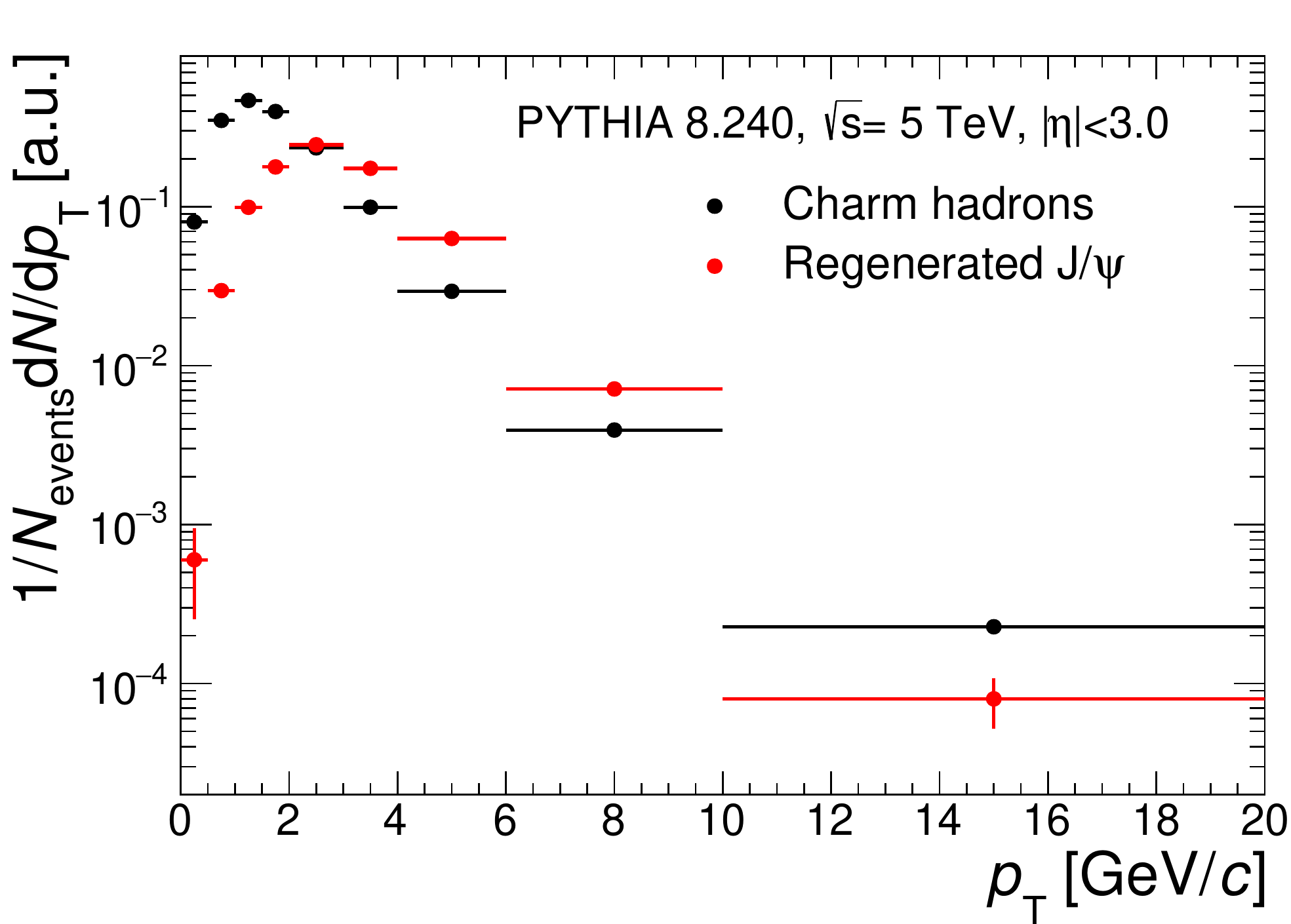}
\includegraphics[width=1.0\columnwidth]{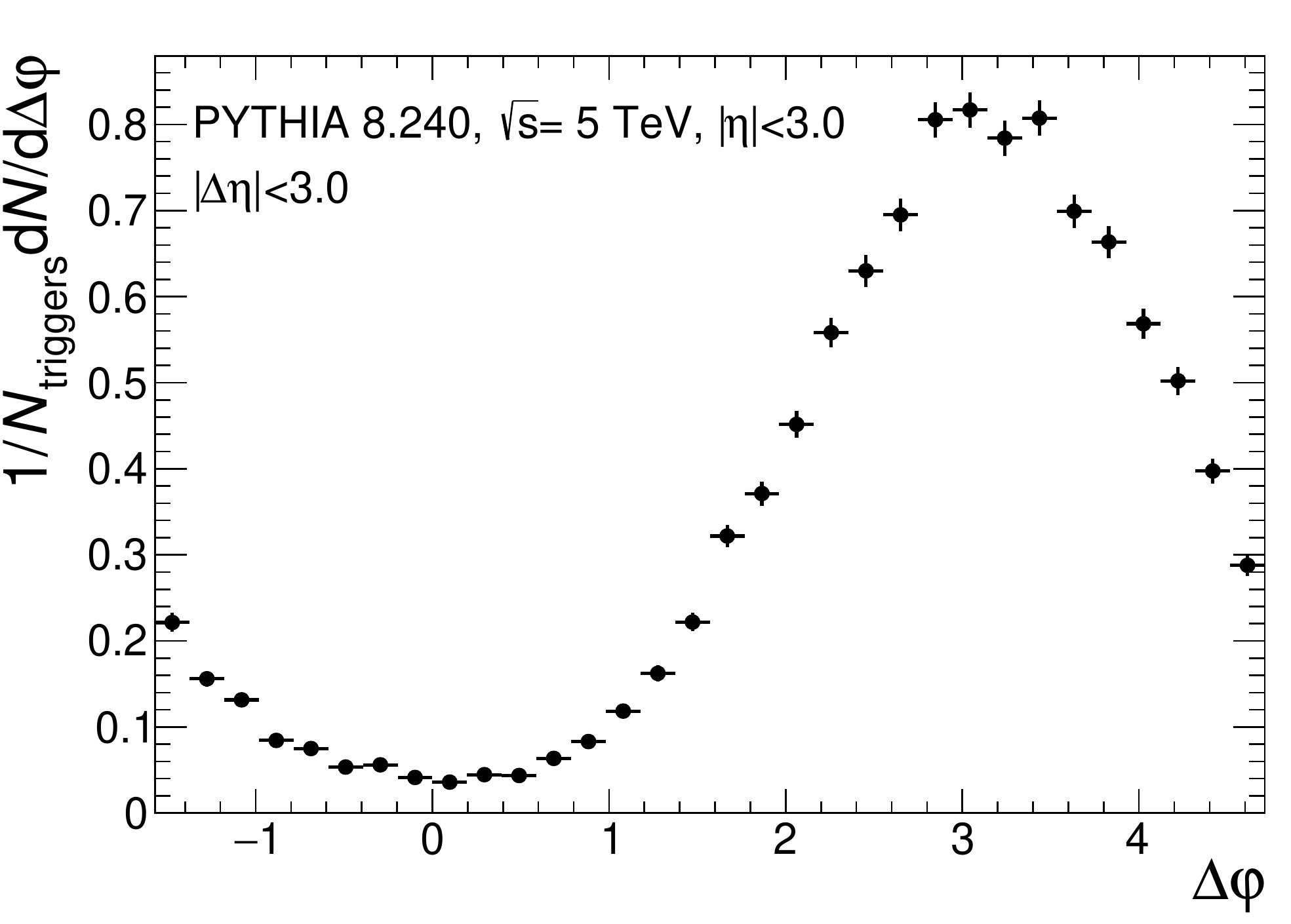}
\caption{Predictions for regenerated \jpsi hadrons based on PYTHIA
  simulations, see text for details. Top: the comparison between the \pt spectra for charmed ``quarks'' and regenerated \jpsi. Note that the spectra have been scaled to have a similar magnitude so the shape is easy to compare. Bottom: azimuthal correlations between regenerated \jpsi and their two correlated charm-hadron
  partners.}
\label{fig:jpsi_corr_pythia}
\end{figure}

To estimate how these correlations could look, we have utilized the PYTHIA
simulations discussed in the previous section. We have matched charm
``quarks''\footnote{We used charm hadrons as proxies for the quarks.} from two
different \pp (nucleon-nucleon) collisions by requiring that their rapidity,
azimuthal angle and \pt are close. In the case that the two ``quarks'' are
close we have combined them into a \jpsi with similar angle and rapidity and
approximately twice the \pt. Then we have measured the angular correlations
with respect to the remaining ``associated'' charm hadrons in the
event. Figure~\ref{fig:jpsi_corr_pythia} shows the \pt spectrum of the
regenerated \jpsi (top) and the azimuthal correlation between the \jpsi and
the ``associated'' charm hadrons (bottom). As can be seen, there is a strong
correlation on the away side. The large magnitude is easy to understand, as
each regenerated \jpsi will have two charm hadron partners, and that it is
primarily on the away side is easy to understand from
Fig.~\ref{fig:jpsi_corr_pythia}. Due to possible NLO corrections discussed
previously, an experimentally observed balance will not necessarily be on the
away-side peak only. However, we note that the expected shape is in principle
derivable from the measured charm spectra and charm-anticharm balance
function, discussed in Sec.~\ref{sec:charm:anticharm}. We therefore think that
these measurements can provide clear and unique signatures of \jpsi
regeneration. The only exception to this would be if charm quarks fully
kinematically (but not chemically) thermalize as discussed in
Ref.~\cite{Braun-Munzinger:2000csl}. But again, this would be observable in
measurements of the charm balance function, which in this case would have no
structure. We note that we consider such a scenario very unlikely as what
makes the GP/QGP liquid perfect is exactly that the effect of dissipation and
diffusion is as low as possible. \\

In this paragraph we will go into detail about how we think one could search
for this signature in Runs 3 and 4 at the LHC. To search for the largest signal,
one will have to study central collisions where regeneration is the largest
effect and trigger on low \pt \jpsi where one knows that the regeneration
plays a big role and where the \ccbar pairs are produced back-to-back. An
important crosscheck that a signal is indeed real and from regeneration will
be that one can switch it off both by going to peripheral collisions and by
increasing the trigger \pt. The final thing to consider is what hadrons to
measure. The correlation function in Fig.~\ref{fig:jpsi_corr_pythia} (bottom)
is for all charm hadron species. The important thing will be to try recover
as much as possible of the yield of charm hadrons. \\

In the final part of this section we will go through how we think these
measurements can provide alternative insights. First, we note that the
measurements described here have a strong relation to the balance function
measurements described in Sec.~\ref{sec:charm:anticharm}: in this section we measure how
the regenerated \jpsi is balanced by its original charm partners. Therefore,
the results should be able to provide similar information as discussed in the
previous section.\\ Furthermore, we think that if one would establish this as
a unique signature of \jpsi regeneration then one could even use this
signature to search for similar effects in small collision systems. In this way one
could hope to find evidence of deconfinement even in small systems. \\ We note
that there are ideas similar to \jpsi regeneration even in PYTHIA where it can
occur via a process known as ``cluster
collapse''~\cite{Norrbin:1998bw}. Cluster collapse can occur when ``the string mass is so small that
the cluster cannot decay into two hadrons. It is then assumed to collapse
directly into a hadron resonance, inheriting the flavor content of the string
endpoints'' ,
 (see Ref.~\cite{Norrbin:1998bw} for details). This normally does not produce \jpsi as the charm and anticharm
produced in a single scattering will end up on different strings. However, in
case two or more \ccbar pairs are produced, \jpsi production can occur due to
color reconnection as discussed in detail in Ref.~\cite{Weber:2018ddv}. We would
expect this type of \jpsi production mechanism to have similar signatures as
shown in Fig.~\ref{fig:jpsi_corr} and Fig.~\ref{fig:jpsi_corr_pythia} and note
that this signature, as far as we know, is discussed for the first time here.

One could argue that the PYTHIA implementation raises questions about the
unique interpretation of \jpsi regeneration as a signature of
deconfinement. It should be clear that partons in the GP and the QGP are
strongly interacting so deconfinement does not mean ``free''. We think the
main point here is that in \AAA collisions the magnitude of the \jpsi
regeneration requires that \ccbar pairs from \emph{different} nucleon-nucleon
collisions are required to ``regenerate'', which to us means that the charm
quarks are not confined together with the partons produced locally but can end
up in hadrons with quarks produced in processes that are, at least in essence,
causally separated. For a similar signature in a small system one would not be
able to use this argument and would instead have to determine what is the
most likely scenario: to have one or two production mechanisms.

\section{Conclusions}
\label{sec:conclusions}

In this paper, we have drawn attention to the possible existence of a gluonic
state, which we have denoted the Gluon Plasma (GP), that could dominate
early-time dynamics of hadronic collisions. The dynamics of the GP, which is
critical for understanding the build up of flow, might therefore be invisible
to light quark hadrons as they are first produced \emph{after} the system is
already flowing. We have argued that the evolution of the charm balance
function as one goes from small to large systems would be sensitive to the
early time GP dynamics and therefore charm-charm correlation functions will be
vital measurements in the LHC physics program during Runs 3 and 4. \\ The question of the GP
is fundamentally linked to the question of deconfinement as the GP will
effectively give similar experimental signatures as confined models. For
this reason, we have in the latter part shown exactly how one can combine the
ideas of the charm balance function with \jpsi regeneration to strengthen the
experimental observation of charm quark deconfinement.

\begin{acknowledgements}
PC and AO gratefully acknowledge funding from the Knut and Alice Wallenberg
Foundation (the CLASH project).
\end{acknowledgements}

\bibliographystyle{spphys} \interlinepenalty=10000 \bibliography{reference}

\begin{thebibliography}{10}
\providecommand{\url}[1]{{#1}}
\providecommand{\urlprefix}{URL }
\expandafter\ifx\csname urlstyle\endcsname\relax
  \providecommand{\doi}[1]{DOI \discretionary{}{}{}#1}\else
  \providecommand{\doi}{DOI \discretionary{}{}{}\begingroup
  \urlstyle{rm}\Url}\fi

\bibitem{Adams:2005dq}
J.~Adams, et~al., Nucl. Phys. A \textbf{757}, 102 (2005).
\newblock \doi{10.1016/j.nuclphysa.2005.03.085}

\bibitem{Adcox:2004mh}
K.~Adcox, et~al., Nucl. Phys. A \textbf{757}, 184 (2005).
\newblock \doi{10.1016/j.nuclphysa.2005.03.086}

\bibitem{Busza:2018rrf}
W.~Busza, K.~Rajagopal, W.~van~der Schee, Ann. Rev. Nucl. Part. Sci.
  \textbf{68}, 339 (2018).
\newblock \doi{10.1146/annurev-nucl-101917-020852}

\bibitem{Khachatryan:2010gv}
V.~Khachatryan, et~al., JHEP \textbf{09}, 091 (2010).
\newblock \doi{10.1007/JHEP09(2010)091}

\bibitem{Abelev:2012ola}
B.~Abelev, et~al., Phys. Lett. B \textbf{719}, 29 (2013).
\newblock \doi{10.1016/j.physletb.2013.01.012}

\bibitem{Nagle:2018nvi}
J.L. Nagle, W.A. Zajc, Ann. Rev. Nucl. Part. Sci. \textbf{68}, 211 (2018).
\newblock \doi{10.1146/annurev-nucl-101916-123209}

\bibitem{Romatschke:2017ejr}
P.~Romatschke, U.~Romatschke, \emph{{Relativistic Fluid Dynamics In and Out of
  Equilibrium}}.
\newblock Cambridge Monographs on Mathematical Physics (Cambridge University
  Press, 2019).
\newblock \doi{10.1017/9781108651998}

\bibitem{Kurkela:2018xxd}
A.~Kurkela, A.~Mazeliauskas, Phys. Rev. Lett. \textbf{122}, 142301 (2019).
\newblock \doi{10.1103/PhysRevLett.122.142301}

\bibitem{Shuryak:1992wc}
E.V. Shuryak, Phys. Rev. Lett. \textbf{68}, 3270 (1992).
\newblock \doi{10.1103/PhysRevLett.68.3270}

\bibitem{Stocker:2015nka}
H.~Stocker, et~al., Astron. Nachr. \textbf{336}(8/9), 744 (2015).
\newblock \doi{10.1002/asna.201512252}

\bibitem{Shuryak:2019ydl}
E.~Shuryak, arXiv:1901.00178 [nucl-th]  (2019)

\bibitem{ALICE:2018bdo}
S.~Acharya, et~al., JHEP \textbf{02}, 012 (2019).
\newblock \doi{10.1007/JHEP02(2019)012}

\bibitem{Adare:2009qk}
A.~Adare, et~al., Phys. Rev. C \textbf{81}, 034911 (2010).
\newblock \doi{10.1103/PhysRevC.81.034911}

\bibitem{Adam:2015lda}
J.~Adam, et~al., Phys. Lett. B \textbf{754}, 235 (2016).
\newblock \doi{10.1016/j.physletb.2016.01.020}

\bibitem{Adare:2011zr}
A.~Adare, et~al., Phys. Rev. Lett. \textbf{109}, 122302 (2012).
\newblock \doi{10.1103/PhysRevLett.109.122302}

\bibitem{Liu:2012ax}
F.M. Liu, S.X. Liu, Phys. Rev. C \textbf{89}(3), 034906 (2014).
\newblock \doi{10.1103/PhysRevC.89.034906}

\bibitem{Monnai:2014kqa}
A.~Monnai, Phys. Rev. C \textbf{90}(2), 021901 (2014).
\newblock \doi{10.1103/PhysRevC.90.021901}

\bibitem{Kharzeev:2013ffa}
D.E. Kharzeev, Prog. Part. Nucl. Phys. \textbf{75}, 133 (2014).
\newblock \doi{10.1016/j.ppnp.2014.01.002}

\bibitem{Kharzeev:2015kna}
D.E. Kharzeev, Ann. Rev. Nucl. Part. Sci. \textbf{65}, 193 (2015).
\newblock \doi{10.1146/annurev-nucl-102313-025420}

\bibitem{Li:2014bha}
Q.~Li, D.E. Kharzeev, C.~Zhang, Y.~Huang, I.~Pletikosic, A.V. Fedorov, R.D.
  Zhong, J.A. Schneeloch, G.D. Gu, T.~Valla, Nature Phys. \textbf{12}, 550
  (2016).
\newblock \doi{10.1038/nphys3648}

\bibitem{Acharya:2017fau}
S.~Acharya, et~al., Phys. Lett. B \textbf{777}, 151 (2018).
\newblock \doi{10.1016/j.physletb.2017.12.021}

\bibitem{Khachatryan:2016got}
V.~Khachatryan, et~al., Phys. Rev. Lett. \textbf{118}(12), 122301 (2017).
\newblock \doi{10.1103/PhysRevLett.118.122301}

\bibitem{Acharya:2020rlz}
S.~Acharya, et~al., JHEP \textbf{09}, 160 (2020).
\newblock \doi{10.1007/JHEP09(2020)160}

\bibitem{ALICE:2017sss}
S.~Acharya, et~al., Phys. Lett. B \textbf{777}, 151 (2018).
\newblock \doi{10.1016/j.physletb.2017.12.021}

\bibitem{STAR:2021mii}
M.~Abdallah, et~al., arXiv:2109.00131 [nucl-ex]  (2021)

\bibitem{Bass:2000az}
S.A. Bass, P.~Danielewicz, S.~Pratt, Phys.\ Rev.\ Lett. \textbf{85}, 2689
  (2000).
\newblock \doi{10.1103/PhysRevLett.85.2689}.
\newblock \urlprefix\url{http://link.aps.org/doi/10.1103/PhysRevLett.85.2689}

\bibitem{Nahrgang:2013saa}
M.~Nahrgang, J.~Aichelin, P.B. Gossiaux, K.~Werner, Phys. Rev. C
  \textbf{90}(2), 024907 (2014).
\newblock \doi{10.1103/PhysRevC.90.024907}

\bibitem{Jeon:2001ue}
S.~Jeon, S.~Pratt, Phys.\ Rev. \textbf{C65}, 044902 (2002).
\newblock \doi{10.1103/PhysRevC.65.044902}.
\newblock \urlprefix\url{http://link.aps.org/doi/10.1103/PhysRevC.65.044902}

\bibitem{Pratt:2015jsa}
S.~Pratt, W.P. McCormack, C.~Ratti, Phys.\ Rev.\ {\bf C} \textbf{92}, 064905
  (2015).
\newblock \doi{10.1103/PhysRevC.92.064905}.
\newblock \urlprefix\url{http://link.aps.org/doi/10.1103/PhysRevC.92.064905}

\bibitem{Bialas200431}
A.~Bialas, Phys.\ Lett. \textbf{B579}(1?2), 31  (2004).
\newblock \doi{http://dx.doi.org/10.1016/j.physletb.2003.10.106}.
\newblock
  \urlprefix\url{http://www.sciencedirect.com/science/article/pii/S0370269303017155}

\bibitem{Abelev:2013csa}
B.~Abelev, et~al., Phys. Lett. B \textbf{723}, 267 (2013).
\newblock \doi{10.1016/j.physletb.2013.05.039}

\bibitem{Adam:2015gda}
J.~Adam, et~al., Eur. Phys. J. C \textbf{76}(2), 86 (2016).
\newblock \doi{10.1140/epjc/s10052-016-3915-1}

\bibitem{Andersson:1983ia}
B.~Andersson, G.~Gustafson, G.~Ingelman, T.~Sjostrand, Phys. Rept. \textbf{97},
  31 (1983).
\newblock \doi{10.1016/0370-1573(83)90080-7}

\bibitem{Sjostrand:2007gs}
T.~Sjostrand, S.~Mrenna, P.Z. Skands, Comput. Phys. Commun. \textbf{178}, 852
  (2008).
\newblock \doi{10.1016/j.cpc.2008.01.036}

\bibitem{Sjostrand:2014zea}
T.~Sj\"ostrand, S.~Ask, J.R. Christiansen, R.~Corke, N.~Desai, P.~Ilten,
  S.~Mrenna, S.~Prestel, C.O. Rasmussen, P.Z. Skands, Comput. Phys. Commun.
  \textbf{191}, 159 (2015).
\newblock \doi{10.1016/j.cpc.2015.01.024}

\bibitem{Matsui:1986dk}
T.~Matsui, H.~Satz, Phys. Lett. B \textbf{178}, 416 (1986).
\newblock \doi{10.1016/0370-2693(86)91404-8}

\bibitem{Abreu:2000ni}
M.C. Abreu, et~al., Phys. Lett. B \textbf{477}, 28 (2000).
\newblock \doi{10.1016/S0370-2693(00)00237-9}

\bibitem{Adare:2006ns}
A.~Adare, et~al., Phys. Rev. Lett. \textbf{98}, 232301 (2007).
\newblock \doi{10.1103/PhysRevLett.98.232301}

\bibitem{Dong:2019byy}
X.~Dong, Y.J. Lee, R.~Rapp, Ann. Rev. Nucl. Part. Sci. \textbf{69}, 417 (2019).
\newblock \doi{10.1146/annurev-nucl-101918-023806}

\bibitem{Vogt:2018oje}
R.~Vogt, Phys. Rev. C \textbf{98}(3), 034907 (2018).
\newblock \doi{10.1103/PhysRevC.98.034907}

\bibitem{Akamatsu:2009ya}
Y.~Akamatsu, T.~Hatsuda, T.~Hirano, Phys. Rev. C \textbf{80}, 031901 (2009).
\newblock \doi{10.1103/PhysRevC.80.031901}

\bibitem{Zhu:2006er}
X.~Zhu, M.~Bleicher, S.L. Huang, K.~Schweda, H.~Stoecker, N.~Xu, P.~Zhuang,
  Phys. Lett. B \textbf{647}, 366 (2007).
\newblock \doi{10.1016/j.physletb.2007.01.072}

\bibitem{Tsiledakis:2009qh}
G.~Tsiledakis, H.~Appelshauser, K.~Schweda, J.~Stachel, Nucl. Phys. A
  \textbf{858}, 86 (2011).
\newblock \doi{10.1016/j.nuclphysa.2011.03.013}

\bibitem{Ruggieri:2018ies}
M.~Ruggieri, S.K. Das, EPJ Web Conf. \textbf{192}, 00017 (2018).
\newblock \doi{10.1051/epjconf/201819200017}

\bibitem{Adolfsson:2020dhm}
J.~Adolfsson, et~al., Eur. Phys. J. A \textbf{56}(11), 288 (2020).
\newblock \doi{10.1140/epja/s10050-020-00270-1}

\bibitem{Adolfsson}
J.~Adolfsson, {Study of $\Xi$-Hadron Correlations in pp Collisions at $\sqrt{s}
  = 13$ TeV Using the ALICE Detector}.
\newblock Ph.D. thesis, Lund University (2020).
\newblock \urlprefix\url{http://cds.cern.ch/record/2750097/}

\bibitem{Adolfsson:2020sxz}
J.~Adolfsson, Acta Phys. Polon. Supp. \textbf{14}, 21 (2021).
\newblock \doi{10.5506/APhysPolBSupp.14.21}

\bibitem{Basu:2020ldt}
S.~Basu, V.~Gonzalez, J.~Pan, A.~Knospe, A.~Marin, C.~Markert, C.~Pruneau,
  (2020).
\newblock \urlprefix\url{arXiv 2001.07167 [nucl-ex]}

\bibitem{Sjostrand:2006za}
T.~Sjostrand, S.~Mrenna, P.Z. Skands, JHEP \textbf{05}, 026 (2006).
\newblock \doi{10.1088/1126-6708/2006/05/026}

\bibitem{LHCb:2012aiv}
R.~Aaij, et~al., JHEP \textbf{06}, 141 (2012).
\newblock \doi{10.1007/JHEP06(2012)141}.
\newblock [Addendum: JHEP 03, 108 (2014)]

\bibitem{Schukraft:2012ah}
J.~Schukraft, A.~Timmins, S.A. Voloshin, Phys. Lett. B \textbf{719}, 394
  (2013).
\newblock \doi{10.1016/j.physletb.2013.01.045}

\bibitem{Abelev:2012rv}
B.~Abelev, et~al., Phys. Rev. Lett. \textbf{109}, 072301 (2012).
\newblock \doi{10.1103/PhysRevLett.109.072301}

\bibitem{Abelev:2013ila}
B.B. Abelev, et~al., Phys. Lett. B \textbf{734}, 314 (2014).
\newblock \doi{10.1016/j.physletb.2014.05.064}

\bibitem{BraunMunzinger:2000px}
P.~Braun-Munzinger, J.~Stachel, Phys. Lett. B \textbf{490}, 196 (2000).
\newblock \doi{10.1016/S0370-2693(00)00991-6}

\bibitem{Braun-Munzinger:2000csl}
P.~Braun-Munzinger, J.~Stachel, Phys. Lett. B \textbf{490}, 196 (2000).
\newblock \doi{10.1016/S0370-2693(00)00991-6}

\bibitem{Norrbin:1998bw}
E.~Norrbin, T.~Sjostrand, Phys. Lett. B \textbf{442}, 407 (1998).
\newblock \doi{10.1016/S0370-2693(98)01244-1}

\bibitem{Weber:2018ddv}
S.G. Weber, A.~Dubla, A.~Andronic, A.~Morsch, Eur. Phys. J. C \textbf{79}(1),
  36 (2019).
\newblock \doi{10.1140/epjc/s10052-018-6531-4}

\end{thebibliography}

\end{document}